\documentclass[12pt]{article}
\usepackage{epsfig}
\usepackage{amsmath}
\usepackage{hhline}
\usepackage{amssymb}
\usepackage{amsbsy}
\usepackage{mathptmx}
\usepackage{cite}
\usepackage{lineno}

\newlength{\dinwidth}
\newlength{\dinmargin}
\begin{document}  
% The rest
\newcommand{\pom}{{I\!\!P}}
\newcommand{\reg}{{I\!\!R}}
\newcommand{\slowpi}{\pi_{\mathit{slow}}}
\newcommand{\fiidiii}{F_2^{D(3)}}
\newcommand{\fiidiiiarg}{\fiidiii\,(\beta,\,Q^2,\,x)}
\newcommand{\n}{1.19\pm 0.06 (stat.) \pm0.07 (syst.)}
\newcommand{\nz}{1.30\pm 0.08 (stat.)^{+0.08}_{-0.14} (syst.)}
\newcommand{\fiidiiiful}{F_2^{D(4)}\,(\beta,\,Q^2,\,x,\,t)}
\newcommand{\fiipom}{\tilde F_2^D}
\newcommand{\ALPHA}{1.10\pm0.03 (stat.) \pm0.04 (syst.)}
\newcommand{\ALPHAZ}{1.15\pm0.04 (stat.)^{+0.04}_{-0.07} (syst.)}
\newcommand{\fiipomarg}{\fiipom\,(\beta,\,Q^2)}
\newcommand{\pomflux}{f_{\pom / p}}
\newcommand{\nxpom}{1.19\pm 0.06 (stat.) \pm0.07 (syst.)}
\newcommand {\gapprox}
   {\raisebox{-0.7ex}{$\stackrel {\textstyle>}{\sim}$}}
\newcommand {\lapprox}
   {\raisebox{-0.7ex}{$\stackrel {\textstyle<}{\sim}$}}
\def\gsim{\,\lower.25ex\hbox{$\scriptstyle\sim$}\kern-1.30ex%
\raise 0.55ex\hbox{$\scriptstyle >$}\,}
\def\lsim{\,\lower.25ex\hbox{$\scriptstyle\sim$}\kern-1.30ex%
\raise 0.55ex\hbox{$\scriptstyle <$}\,}
\newcommand{\pomfluxarg}{f_{\pom / p}\,(x_\pom)}
\newcommand{\dsf}{\mbox{$F_2^{D(3)}$}}
\newcommand{\dsfva}{\mbox{$F_2^{D(3)}(\beta,Q^2,x_{I\!\!P})$}}
\newcommand{\dsfvb}{\mbox{$F_2^{D(3)}(\beta,Q^2,x)$}}
\newcommand{\dsfpom}{$F_2^{I\!\!P}$}
\newcommand{\gap}{\stackrel{>}{\sim}}
\newcommand{\lap}{\stackrel{<}{\sim}}
\newcommand{\fem}{$F_2^{em}$}
\newcommand{\tsnmp}{$\tilde{\sigma}_{NC}(e^{\mp})$}
\newcommand{\tsnm}{$\tilde{\sigma}_{NC}(e^-)$}
\newcommand{\tsnp}{$\tilde{\sigma}_{NC}(e^+)$}
\newcommand{\st}{$\star$}
\newcommand{\sst}{$\star \star$}
\newcommand{\ssst}{$\star \star \star$}
\newcommand{\sssst}{$\star \star \star \star$}
\newcommand{\ra}{\rightarrow}
\newcommand{\tw}{\theta_W}
\newcommand{\sw}{\sin{\theta_W}}
\newcommand{\cw}{\cos{\theta_W}}
\newcommand{\sww}{\sin^2{\theta_W}}
\newcommand{\cww}{\cos^2{\theta_W}}
\newcommand{\trm}{m_{\perp}}
\newcommand{\trp}{p_{\perp}}
\newcommand{\trmm}{m_{\perp}^2}
\newcommand{\trpp}{p_{\perp}^2}
\newcommand{\alp}{\alpha_s}

\newcommand{\alps}{\alpha_s}
\newcommand{\sqrts}{$\sqrt{s}$}
\newcommand{\LO}{$O(\alpha_s^0)$}
\newcommand{\Oa}{$O(\alpha_s)$}
\newcommand{\Oaa}{$O(\alpha_s^2)$}
\newcommand{\PT}{p_{\perp}}
\newcommand{\pt}{p_{_{\rm T}}}
\newcommand{\JPSI}{J/\psi}
\newcommand{\sh}{\hat{s}}
\newcommand{\uh}{\hat{u}}
\newcommand{\MP}{m_{J/\psi}}
\newcommand{\PO}{I\!\!P}
\newcommand{\xbj}{x}
\newcommand{\xpom}{x_{\PO}}
\newcommand{\ttbs}{\char'134}
\newcommand{\xpomlo}{3\times10^{-4}}  
\newcommand{\xpomup}{0.05}  
\newcommand{\dgr}{^\circ}
\newcommand{\pbarnt}{\,\mbox{{\rm pb$^{-1}$}}}
\newcommand{\gev}{\,\mbox{GeV}}
\newcommand{\mev}{\,\mbox{MeV}}
\newcommand{\WBoson}{\mbox{$W$}}
\newcommand{\fbarn}{\,\mbox{{\rm fb}}}
\newcommand{\fbarnt}{\,\mbox{{\rm fb$^{-1}$}}}

\newcommand{\GeV}{\; \rm GeV}
\newcommand{\GeVSq}{\; \rm GeV^2}
\newcommand{\MeV}{\, \rm MeV}
\newcommand{\TeV}{\, \rm TeV}
\newcommand{\pb}{$\, \rm pb$~}
\newcommand{\cm}{\rm cm}
\newcommand{\hdick}{\noalign{\hrule height1.4pt}}

\newcommand{\dedxf}{{\rm d}E/{\rm d}x}
\newcommand{\dedx}{${\rm d}E/{\rm d}x$~}
\newcommand{\dedxns}{${\rm d}E/{\rm d}x$}
\newcommand{\ks}{{$K^0_s$~}}
\newcommand{\ksf}{K^0_s}
\newcommand{\knullf}{K^0}
\newcommand{\knull}{$K^0$~}
\newcommand{\thpl}{$\Theta^+$~}
\newcommand{\thplns}{$\Theta^+$}
\newcommand{\thplf}{\Theta^+}
\newcommand{\sigmath}{$\sigma^{\Theta}$}
\newcommand{\sigmathf}{\sigma^{\Theta}}

\newcommand{\sigmaulth}{$\sigma_{UL}^{\Theta}$}
\newcommand{\sigmaulthf}{\sigma_{UL}^{\Theta}}

\newcommand{\sigmaulthpl}{$\sigma_{UL}^{\Theta^+}$}
\newcommand{\sigmaulthplf}{\sigma_{UL}^{\Theta^+}}

\newcommand{\sigmaulthmi}{$\sigma_{UL}^{\overline{\Theta^+}}$}
\newcommand{\sigmaulthmif}{\sigma_{UL}^{\overline{\Theta^+}}}

\newcommand{\thf}{\Theta}
\newcommand{\thmi}{$\overline{\Theta^+}$~}
\newcommand{\thmif}{\overline{\Theta^+}}
\newcommand{\thplxs}{$\sigma(ep\ra e\thplf X \ra e \knullf p  X)$}
\newcommand{\thplxsf}{\sigma(ep\ra e\thplf X \ra e \knullf p  X)}

%	
% Some useful tex commands
%
\newcommand{\qsq}{\ensuremath{Q^2} }
\newcommand{\gevsq}{\ensuremath{\mathrm{GeV}^2} }
\newcommand{\et}{\ensuremath{E_t^*} }
\newcommand{\rap}{\ensuremath{\eta^*} }
\newcommand{\gp}{\ensuremath{\gamma^*}p }
\newcommand{\dsiget}{\ensuremath{{\rm d}\sigma_{ep}/{\rm d}E_t^*} }
\newcommand{\dsigrap}{\ensuremath{{\rm d}\sigma_{ep}/{\rm d}\eta^*} }
% Journal macro
\def\Journal#1#2#3#4{{#1} {\bf #2} (#3) #4}
\def\NCA{\em Nuovo Cimento}
\def\NIM{\em Nucl. Instrum. Methods}
\def\NIMA{{\em Nucl. Instrum. Methods} {\bf A}}
\def\NPB{{\em Nucl. Phys.}   {\bf B}}
\def\PLB{{\em Phys. Lett.}   {\bf B}}
\def\PRL{\em Phys. Rev. Lett.}
\def\PRD{{\em Phys. Rev.}    {\bf D}}
\def\ZPC{{\em Z. Phys.}      {\bf C}}
\def\EJC{{\em Eur. Phys. J.} {\bf C}}
\def\CPC{\em Comp. Phys. Commun.}
\begin{titlepage}

\noindent
\begin{flushleft}
DESY 06-044\hfill ISSN 0418-9833\\
April 2006
\end{flushleft}

\vspace{2cm}

\begin{center}
\begin{Large}

{\bf Search for a Narrow Baryonic Resonance \\
    Decaying to  {\boldmath $K^0_sp$} or {\boldmath $K^0_s\bar p$}\\
    in Deep Inelastic Scattering at HERA}

\vspace{2cm}

H1 Collaboration

\end{Large}
\end{center}

\vspace{2cm}

\begin{abstract}
A search for a narrow baryonic 
resonance decaying to $K^0_sp$ or $K^0_s\bar p$ 
is carried out in deep inelastic $ep$ scattering with
the H1 detector at HERA. Such a resonance could be a strange pentaquark
\thplns, evidence for which has been reported by 
several experiments.
The  $K^0_sp$ and $K^0_s\bar p$  invariant mass distributions presented here 
do not show any significant peak 
in the mass range from threshold up to $1.7\GeV$. 
Mass dependent upper limits on 
$\sigma(ep\ra e\thplf X )\times BR(\thplf \ra  \knullf p)$
are obtained at the 95\% confidence level.
\end{abstract}

\vspace{1.5cm}

\begin{center}
Submitted to \PLB
\end{center}

\end{titlepage}

\begin{flushleft}
  %-- H1AUTS Author list by names 
%-- Status: Mon Mar  6 16:07:34 CET 2006  Number of authors = 295 

A.~Aktas$^{9}$,                %DESY-PD        09/05           Aktas               
V.~Andreev$^{25}$,             %LPI -PD        8/88            Andreev             
T.~Anthonis$^{3}$,             %ANTW-ST        11/99           Anthonis            
B.~Antunovic$^{26}$,           %MPIM-ST        09/03           Antunovic           
S.~Aplin$^{9}$,                %DESY-PD        01/04           Aplin               
A.~Asmone$^{33}$,              %ROME-ST        07/2            Asmone              
A.~Astvatsatourov$^{3}$,       %BRUX-PD        07/04           Astvatsatourov      
A.~Babaev$^{24, \dagger}$,     %ITEP-LEFT      12/05           Babaev              
S.~Backovic$^{30}$,            %PODG-PD        03/2            Backovic            
A.~Baghdasaryan$^{37}$
% ...

%-- H1 Institutes 
\bigskip{\it
% ...
$ ^{3}$ Inter-University Institute for High Energies ULB-VUB, Brussels;
          Universiteit Antwerpen, Antwerpen; Belgium$^{ c}$ \\
% ...
$ ^{9}$ DESY, Hamburg, Germany \\
% ...
$ ^{24}$ Institute for Theoretical and Experimental Physics,
          Moscow, Russia$^{ k}$ \\
 $ ^{25}$ Lebedev Physical Institute, Moscow, Russia$^{ e}$ \\
 $ ^{26}$ Max-Planck-Institut f\"ur Physik, M\"unchen, Germany \\
% ...
$ ^{30}$ Faculty of Science, University of Montenegro,
          Podgorica, Serbia and Montenegro$^{ e}$ \\
% ...
$ ^{33}$ Dipartimento di Fisica Universit\`a di Roma Tre
          and INFN Roma~3, Roma, Italy \\
% ...
$ ^{37}$ Yerevan Physics Institute, Yerevan, Armenia \\
% ...
$ ^{\dagger}$ Deceased \\

\bigskip
% ...
$ ^c$ Supported by FNRS-FWO-Vlaanderen, IISN-IIKW and IWT
      and  by Interuniversity
Attraction Poles Programme,
      Belgian Science Policy \\
% ...
$ ^e$ Supported by the Deutsche Forschungsgemeinschaft \\
% ...
$ ^k$ Partially Supported by Russian Foundation
      for Basic Research,  grants  03-02-17291
      and  04-02-16445 \\
}
\end{flushleft}

\newpage

\section{Introduction}
Recently several fixed-target experiments have published evidence 
for the production
of a strange pentaquark\footnote{In this
paper particle names are used to refer to both the particle 
and its antiparticle,
unless explicitly stated otherwise.} $\thplf$\cite{kenhicks}, 
a hypothetical baryon\cite{theory} 
with a minimal quark content of $uudd\bar s$,
observed in the decay channels
$K^+n$ and $K^0_sp$. 
This state has been reported with masses
in the range of 1520 to 1540$\MeV$ and with a narrow width, 
consistent with the experimental resolution
in most of the observations.
Evidence for \thpl production  has been also obtained 
in deep inelastic $ep$ scattering (DIS) 
at HERA by the ZEUS experiment\cite{zeusspq}.
Many non-observations have also been reported \cite{kenhicks}.
The experimental situation is thus controversial and
further data are needed to establish the existence of 
this resonance.

This paper presents a
search for the strange pentaquark $\thplf$
using $74 \, {\rm pb^{-1}}$ of deep inelastic $ep$ scattering
data taken with the H1 detector in the years 1996-2000.
A narrow resonance is searched for in the $K^0_sp$ or 
$K^0_s\bar p$ decay channel in the mass range from 1.48 to 1.7$\GeV$
and in the kinematic range of negative four momentum transfer squared, $Q^2$, from
5 to 100$\GeVSq$ and of inelasticity, $y$, from 0.1 to 0.6.
\section{Experimental Procedure}
\label{method}
\subsection{H1 Apparatus}
\label{detector}
A detailed description of the H1 detector can be found in \cite{h1det}. 
The following briefly describes only those detector components important 
for the present analysis.

The tracks from charged particles used in this analysis are
reconstructed in the 
central tracker, whose main components are two cylindrical drift chambers,
the inner and outer central jet chamber (CJCs).
The inner and outer CJC are mounted concentrically around the beam-line,
covering the range of pseudorapidities\footnote{
The pseudorapidity is given by $\eta = - \ln \tan \theta / 2$,
where the polar angle $\theta$ is measured with respect to 
the \mbox{$z$ axis} given  by the proton beam direction.}
\mbox{$-1.9 < \eta<1.9$} for tracks coming from the 
nominal event vertex.
The CJCs lie within a homogeneous magnetic field of 
$1.15 \, {\rm T}$ which allows the transverse momentum, $\pt$, 
of charged particles to be measured.
Two additional drift chambers complement the CJCs by precisely
measuring the $z$ coordinates of track segments and hence assist in the
determination of polar angles. 
Two cylindrical multi-wire proportional chambers facilitate 
triggering on tracks.
The transverse momentum resolution of the central tracker is
$\sigma(\pt) / \pt \simeq 0.005 \, \pt \, /\GeV \, \oplus 0.015$.
Charge misidentification is negligible
for particles originating from the primary vertex and having 
transverse momenta in the range relevant to this analysis. 

The specific ionisation energy loss of charged particles, \dedxns, is derived 
from the mean of the inverse square-root
of the charge collected by all CJC sense wires with a signal above a certain
threshold. The average resolution for minimum ionising particles is
$\sigma(\dedxf) / (\dedxf) \simeq 8\%$ \cite{steinhart}.

A lead/scintillating-fibre calorimeter (SpaCal) 
is located in the direction of the
electron beam and covers the pseudorapidity range
$-1.39 < \eta < -3.64$.
It contains  electromagnetic and hadronic sections and
is used here to detect the scattered electron in DIS events and to
measure its energy.
A planar drift chamber, positioned in front of the SpaCal, 
measures the polar angle of the scattered electron track originating
from the event vertex.
The global properties of the hadronic final state are reconstructed 
combining information from the
central tracker, the SpaCal and the Liquid Argon  
calorimeter, which surrounds the central tracker. 
The DIS events studied in this paper are triggered 
on the basis of an energy deposition detected in the SpaCal,
complemented by signals in the CJCs and the multi-wire proportional 
chambers. 

The luminosity measurement is based on the Bethe-Heitler process 
$ep \rightarrow ep\gamma$, where the photon is detected in a 
calorimeter located downstream of the interaction point.
\subsection{Monte Carlo Simulation of \thpl}
\label{models}
To estimate the acceptance for the detection of a hypothetical \thpl state,
a Monte Carlo simulation based on the RAPGAP 3.1 \cite{rapgap}
event generator is used, incorporating 
the Lund string model fragmentation \cite{lund} as implemented in PYTHIA 6.2 \cite{pythia}.
The kinematic distributions of strange baryons in DIS data are 
reasonably well described\cite{strangenessdis} by RAPGAP.
The $\thplf$ is introduced by changing the mass of the 
$\Sigma^{*+}$ to values in the range from 1.48 to 1.7$\GeV$ and
forcing it to decay to $\ksf p$. 
By doing so, it is assumed that the \thpl is produced at pseudorapidities 
and transverse momenta similar to those of 
other strange baryons and that it decays isotropically.
In this simulation the \thpl particle is produced on mass shell.
The generated events are passed through the H1 detector simulation based on
GEANT\cite{geant} and are then subjected to the same 
reconstruction and analysis chain as are the data. 

\subsection{Selection of DIS Events}
The analysis is carried out using data corresponding to an integrated 
luminosity of ${\cal L}=74 \, {\rm pb^{-1}}$, taken in the years
1996-2000. During this time HERA collided 
electrons\footnote{The analysis uses data from periods when
the beam lepton was either a positron (${\cal L}=65\, {\rm pb^{-1}}$) or an electron 
(${\cal L}=9\, {\rm pb^{-1}}$).}  
at an energy of $27.6\GeV$ with protons at 
$820\GeV$ (1996-1997) and $920\GeV$ (1998-2000)\footnote{The
sample with a proton energy of $820 \, (920)\GeV$ corresponds to 
a luminosity of ${\cal L}=18 \, (56) \, {\rm pb^{-1}}$, resulting
in an effective $\sqrt{s}=314\GeV$ for the total sample.}.

Events are selected if the
$z$ coordinate of the event vertex, reconstructed using the 
central tracker, lies within $35 \, {\rm cm}$ of
the mean position for $ep$ interactions. 
The scattered electron is required to be reconstructed in the SpaCal
with an energy, $E_e$, above $11\GeV$.
The negative four momentum transfer squared of the exchanged virtual 
photon, $Q^2$, is required to lie in the range 
$5 < Q^2 < 100\GeVSq$, as
reconstructed from the energy and polar angle of the scattered electron.  
The \mbox{inelasticity $y$} of the event is reconstructed using
the scattered electron kinematics and is required to be in the range
$0.1 < y < 0.6$.
The lower cut on $y$ ensures that the hadronic final state lies 
in the central region of the detector, whilst the upper cut
corresponds approximately to the cut on $E_e$.
The difference between the total energy $E$ and the longitudinal 
component of the total momentum $p_z$, calculated 
from the electron and the hadronic final state, is restricted
to $35 < E - p_z < 70{\GeV}$. This requirement suppresses  
photoproduction background, in which the electron escapes detection and 
a hadron fakes the electron signature.
\subsection{Selection of {\boldmath $K^0_s$} Meson and Proton Candidates}
\label{k0prec}
%%%%%%%%%%%%%%%%%%%%%%%%%%%%%%%%%%%%%%%%%%%%%%%%%%%%%%%%%%%%%%%
%%%  K0s reconstruction  - changed acc. to Dimas selection!
%%%%%%%%%%%%%%%%%%%%%%%%%%%%%%%%%%%%%%%%%%%%%%%%%%%%%%%%%%%%%%%
%
The analysis is based on charged particles reconstructed in the 
central tracker. Tracks are accepted if they 
have transverse momenta $\pt > 0.15\GeV$ and pseudorapidities 
$|\eta|<1.75$.
The \ks meson is identified through its decay into charged pions,
$K^0_s\ra \pi^+\pi^-$. 
Events are accepted if they contain at least one \ks 
candidate 
and at least one proton candidate track
originating from the primary vertex.   

\begin{figure}[bthp]
    \epsfig{figure=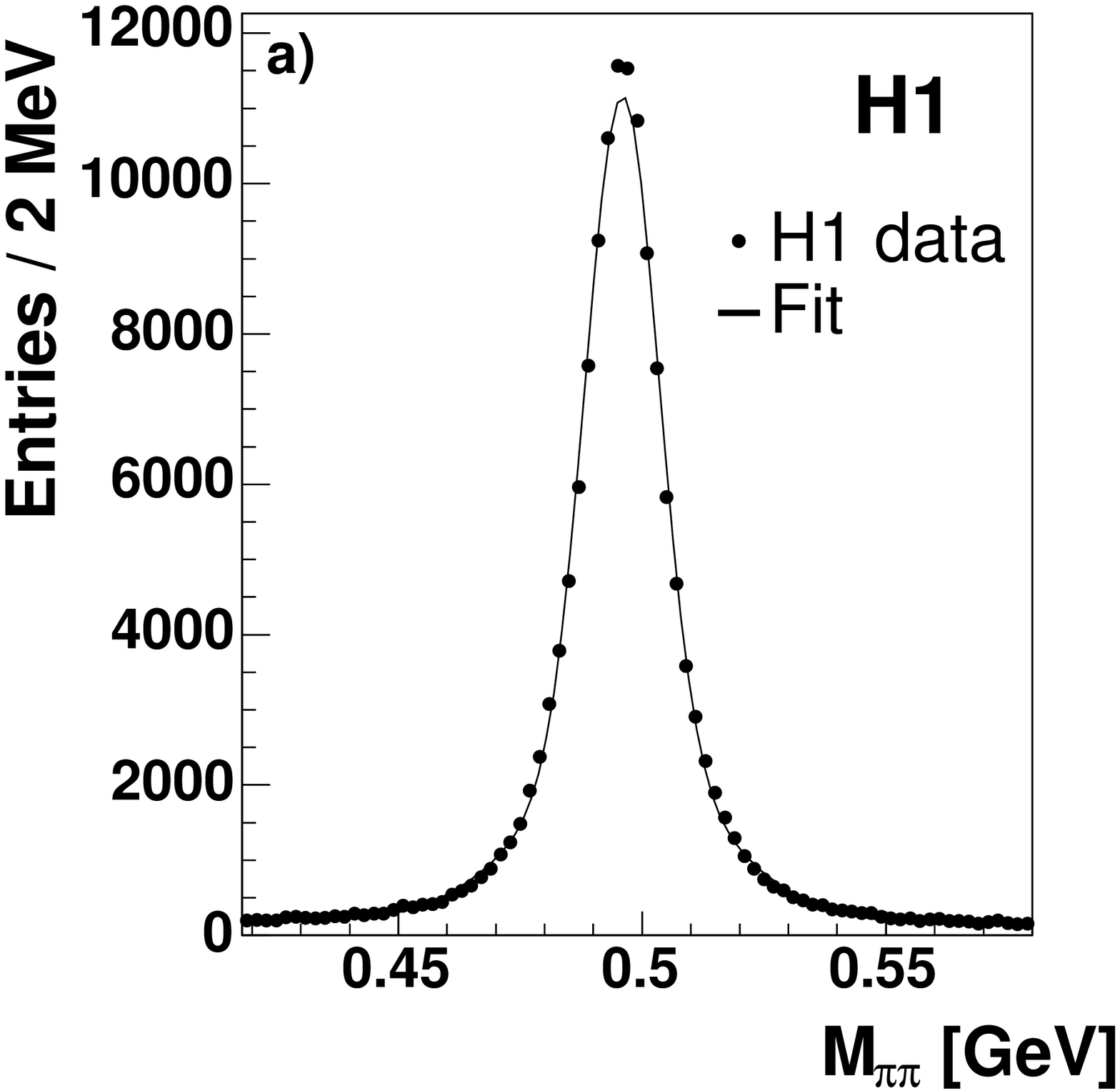,
        width=0.5\textwidth}
    \epsfig{figure=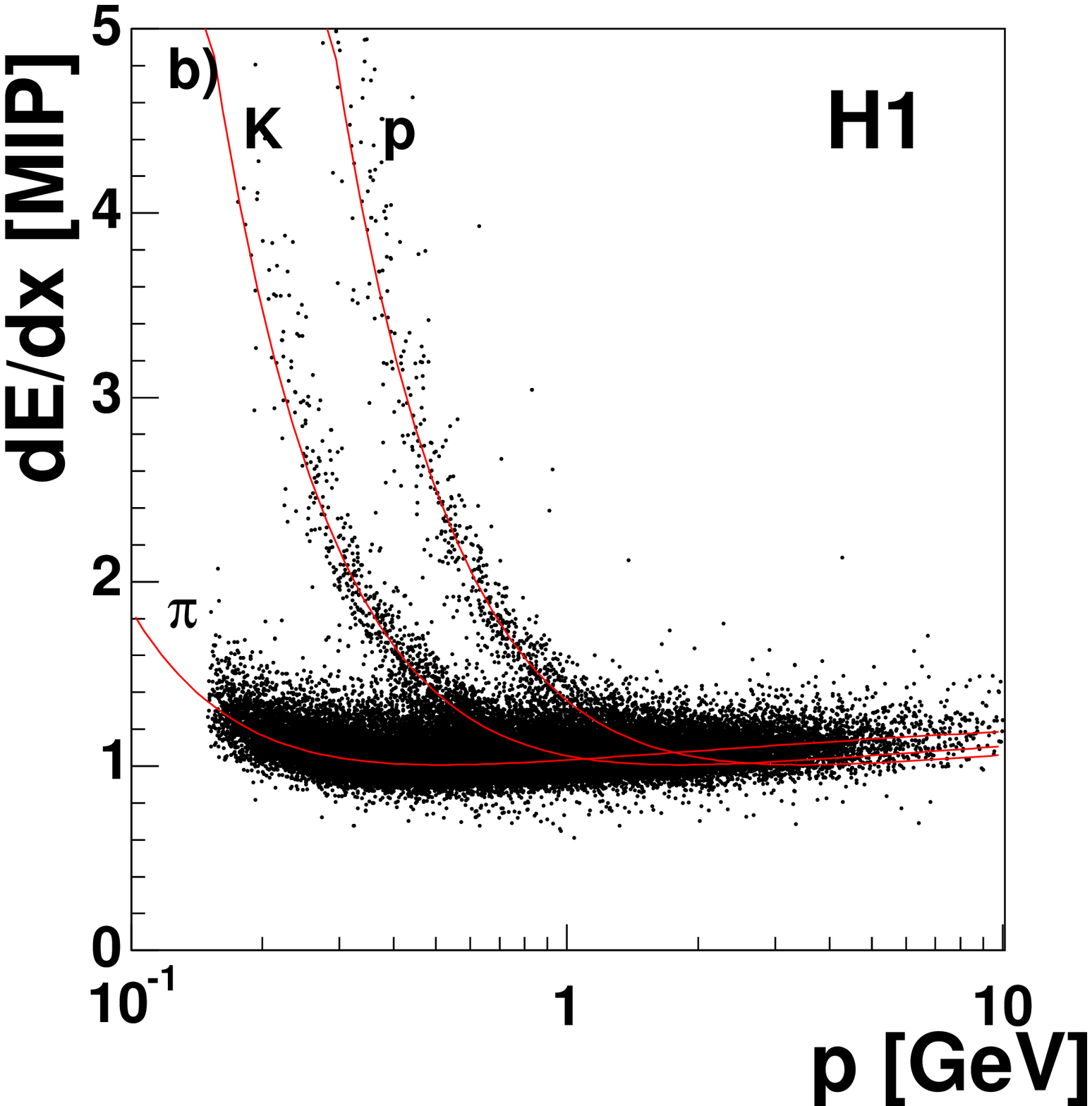,
        width=0.5\textwidth}
    \caption{a) Inclusive $K^0_s$ signal in the invariant $\pi^+\pi^-$ 
             mass distribution for \mbox{$5 < Q^2 < 100\GeVSq$}
	     together with the result from a 
	     fit of a sum of two Gaussian functions for the signal and a 
	     straight line for the background.
             b) Specific  ionisation energy
             loss relative to that of a minimally ionising particle 
             plotted as a function of momentum. 
	     The lines indicate parameterisations
             of the most probable energy loss for pions, kaons and protons
             measured in the CJCs.
     \label{figk0s}}
\end{figure}

\ks candidates are searched for by performing a constrained fit to each pair 
of oppositely charged tracks. The fit demands  these tracks to originate
at a common secondary decay vertex and the decaying neutral particle
to come from the primary vertex.
The secondary vertex must be radially displaced by at least 2 cm from
the primary interaction point. The candidates are required to have a 
transverse momentum
\mbox{$\pt(K^0_s)\ge 0.3\GeV$} and a pseudorapidity
\mbox{$|\eta(K^0_s)|\le 1.5$}.
Contamination from $\Lambda$  production is eliminated
by requiring that the invariant mass
\mbox{$M_{p\pi}$} of the two tracks, reconstructed assigning the
proton (pion) mass to the track with higher (lower) momentum, 
be above  $1.125\GeV$.
Background from converted photons is rejected by the 
requirement \mbox{$M_{ee}> 50\MeV$}.
%%%%%%%%%%%%%%%%%%%%%%%%%%%%%%%%%%%%%%%%%%%%%%%%%%%%%%%%%%%%%%%
% K0s signal 
%%%%%%%%%%%%%%%%%%%%%%%%%%%%%%%%%%%%%%%%%%%%%%%%%%%%%%%%%%%%%%%
%
Figure~\ref{figk0s}a shows the distribution of the invariant mass $M_{\pi^+\pi^-}$ of the 
\ks candidates together with a fit to the data using a
superposition of two Gaussian  functions (to
account for different invariant mass resolutions in
different decay topologies)
and a straight line to approximate the background.
The fitted peak position is 
$M_{\pi^+\pi^-}= 495.9\MeV$ 
which agrees with the nominal \ks mass \cite{pdg}
within a few per mill.
133,000 \ks candidates are reconstructed, as given by subtracting
the fitted background from the data.
\ks candidates with $0.475 < M_{\pi^+\pi^-} < 0.515\GeV$
are selected for further analysis. In this 
mass range the background under the \ks peak is  $\sim 3$\%.

%%%%%%%%%%%%%%%%%%%%%%%%%%%%%%%%%%%%%%%%%%%%%%%%%%%%%%%%%%%%%%%
%% K0s signal
%%%%%%%%%%%%%%%%%%%%%%%%%%%%%%%%%%%%%%%%%%%%%%%%%%%%%%%%%%%%%%%%%%%%%%%%%%%
%%%%%%%%%%%%%%%%%%%%%%%%%%%%%%%%%%%%%%%%%%%%%%%%%%%%%%%%%%%%%%%
%%% proton selection
%%%%%%%%%%%%%%%%%%%%%%%%%%%%%%%%%%%%%%%%%%%%%%%%%%%%%%%%%%%%%%%
Proton candidates are selected using requirements on the 
specific ionisation energy loss, \dedx, measured 
in the CJCs.
Figure~\ref{figk0s}b shows the measured \dedx plotted against
momentum for all tracks originating from the primary vertex, 
which lead to a mass $M_{\ksf p}< 1.8\GeV$
when combined with  the \ks candidates.   The curves in
Fig.~\ref{figk0s}b represent the most probable \dedx values as derived
from a phenomenological parameterisation \cite{steinhart}
based on the Bethe-Bloch formula.
%%%%%%%%%%%%%%%%%%%%%%%%%%%%%%%%%%%%%%%%%%%%%%%%%%%%%%%%%%%%%%%
%%% dE/dx particle Identification
%%%%%%%%%%%%%%%%%%%%%%%%%%%%%%%%%%%%%%%%%%%%%%%%%%%%%%%%%%%%%%%
%%
Likelihoods for a particle to be a pion, kaon or proton
are obtained from the difference between the
measured \dedx and the most probable value for each particle type
at the reconstructed momentum. 
The normalised proton likelihood, $L_p$, is defined as the ratio of the
proton likelihood to the sum of the pion, kaon
and proton likelihoods. 
In order to optimise simultaneously the background suppression 
and the proton selection efficiency, 
a momentum dependent cut on 
the normalised proton likelihood $L_p$ is applied of
$L_p>0.3$ ($L_p>0.1$) for proton momenta below (above) $2\GeV$.
The efficiency of the \dedx selection is tested using protons 
from $\Lambda$ decays.  The efficiency varies  between
65\% and 100\% as a function of momentum and is described by the 
Monte Carlo simulation to within 5\%.

%%%%%%%%%%%%%%%%%%%%%%%%%%%%%%%%%%%%%%%%%%%%%%%%%%%%%%%%%%%%%%%%%%%%%%%%%%%%%%%
\section{Analysis of {\boldmath $\ksf p$} Combinations}
\label{ksp}
\begin{figure}[tbp]
\epsfig{figure=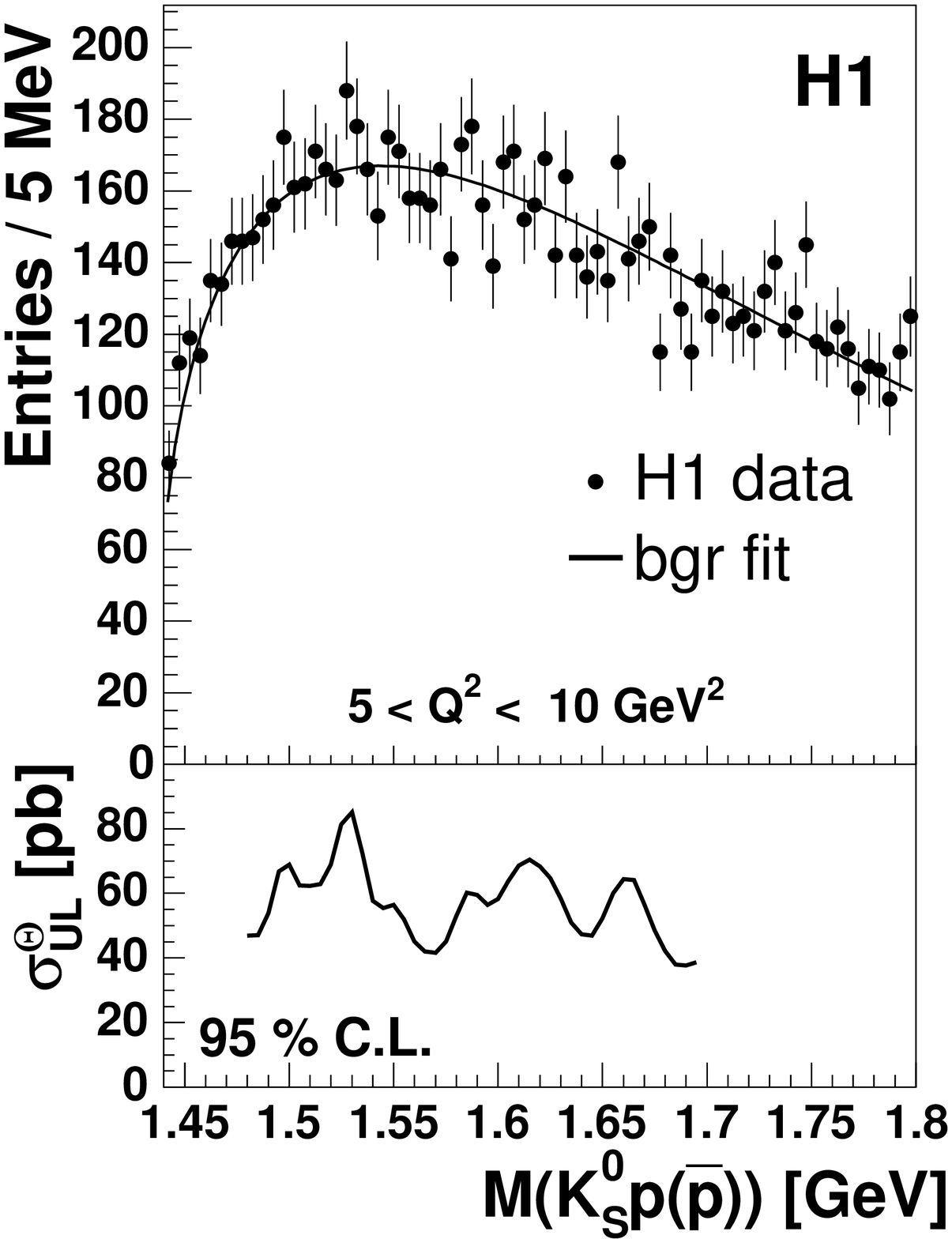,width=0.5\textwidth}
\epsfig{figure=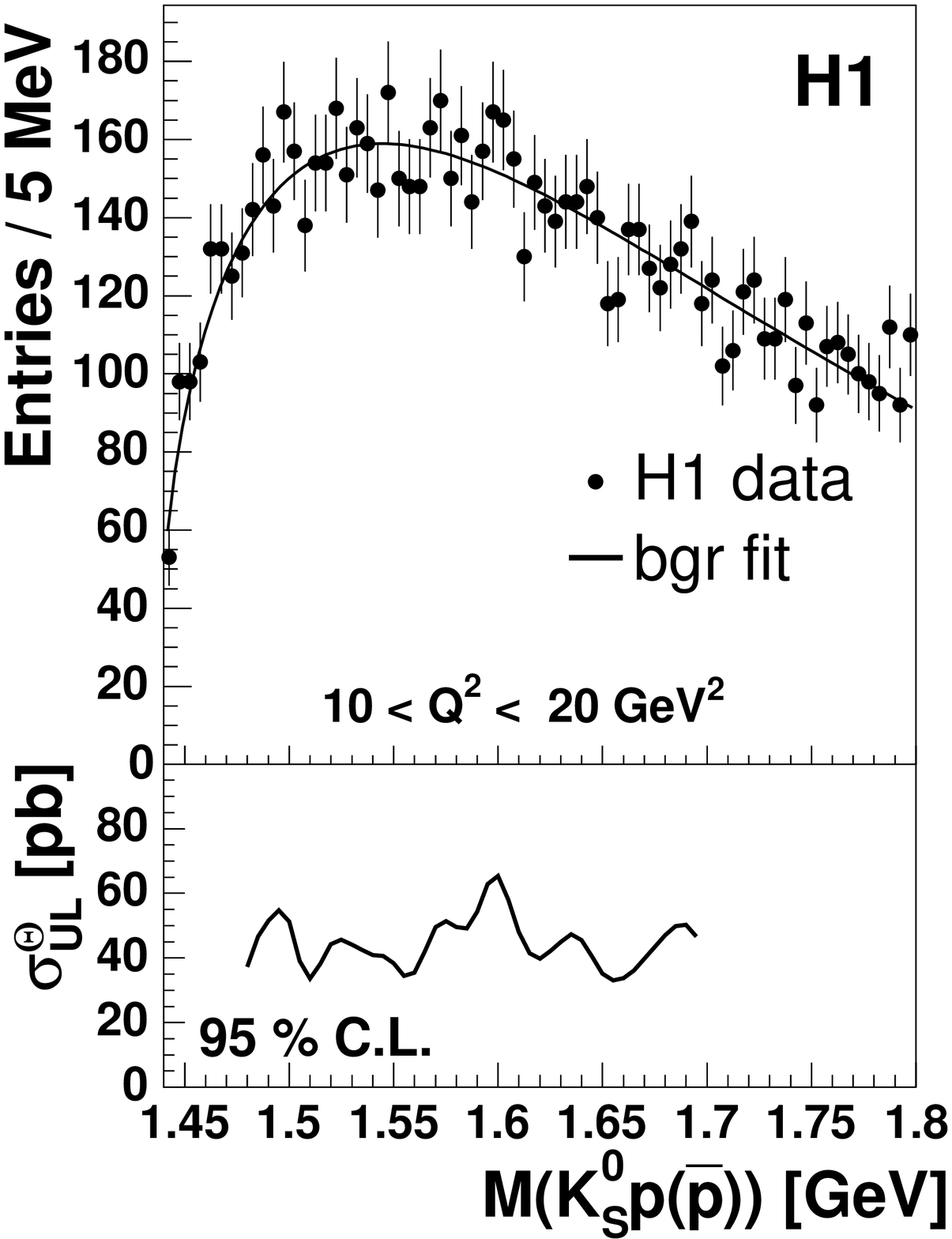,width=0.5\textwidth}
\epsfig{figure=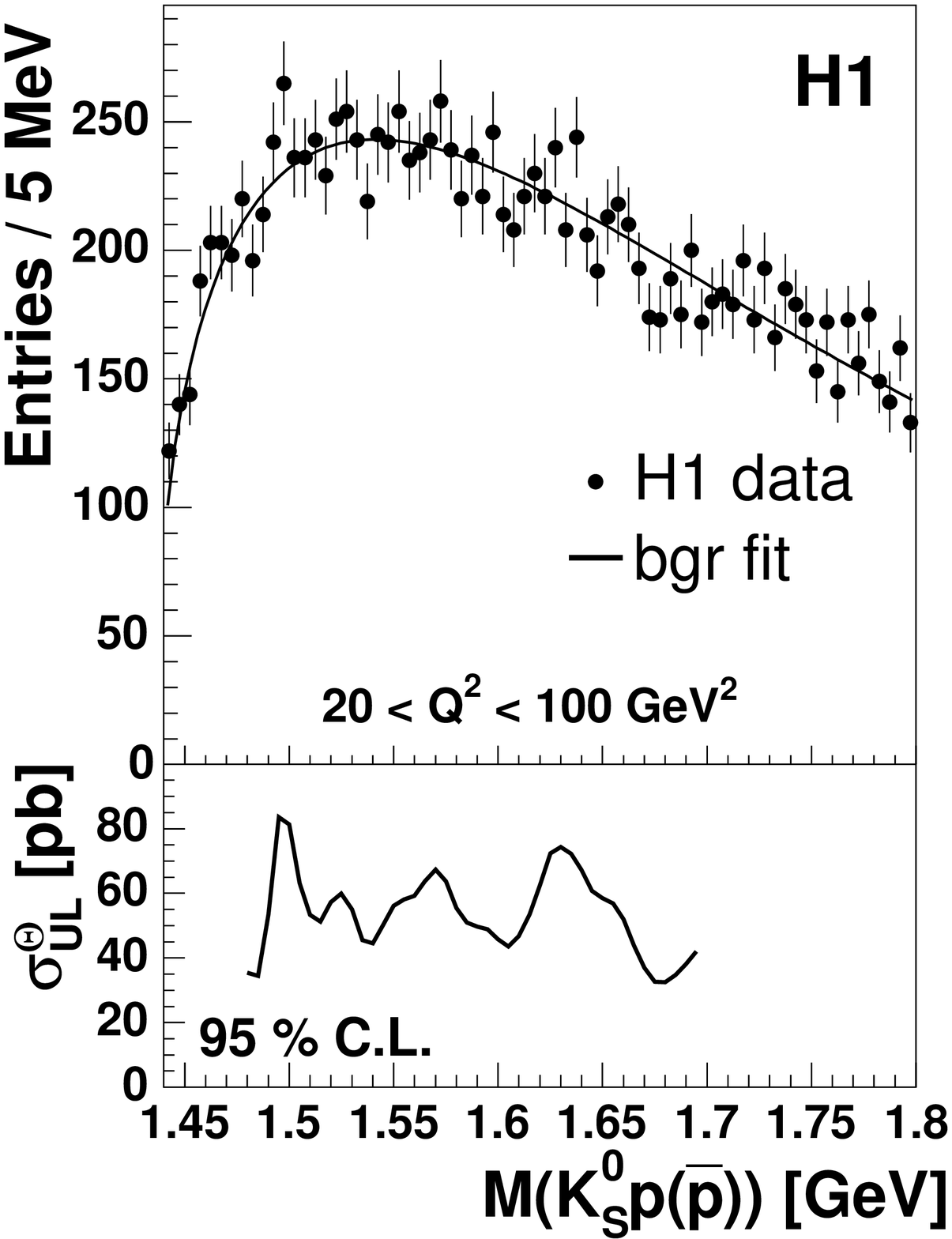,width=0.5\textwidth}
\caption{
Invariant $K^0_s p(\bar p)$ mass spectra in bins of $Q^2$. 
The full line shows the result from the fit of the 
background function (\ref{bgrfit}) to the data. 
The upper limits on the cross section $\sigmaulthf$ (see text)
at 95\% confidence level integrated over the  kinematic range
$\pt(K^0_s p)>0.5\GeV$, $|\eta(K^0_s p)|<1.5$ and $0.1 < y <0.6$
are shown below the mass spectra.
\label{figmkspandlimits}}
\end{figure}

%%%%%%%%%%%%%%%%%%%%%%%%%%%%%%%%%%%%%%%%%%%%%%%%%%%%%%%%%%%%%%%%%%%%%%%%%%%%%%%
%% K0s proton invariant mass
%%%%%%%%%%%%%%%%%%%%%%%%%%%%%%%%%%%%%%%%%%%%%%%%%%%%%%%%%%%%%%%%%%%%%%%%%%
In order to search for a \thpl resonance, 
the candidate \ks mesons are combined with the proton candidates.
To improve the mass resolution the $\ksf p$ four vector is calculated as 
the sum of the \ks and proton
four vectors with 
\mbox{$E_{\ksf}=\sqrt{p_{\ksf}^2 + M_{\ksf}^2}$},
% \mbox{$E^2_{\ksf}={p_{\ksf}^2 + M_{\ksf}^2}$},
where the nominal mass $M_{\ksf}$ is used instead of $M_{\pi^+\pi^-}$.
For the $\ksf p$ system, \mbox{$\pt(\ksf p) >  0.5\GeV$} and 
\mbox{$|\eta(\ksf p) | <  1.5$} are required. 
The $M_{\ksf p}$ distributions are shown in Fig.~\ref{figmkspandlimits}
for three bins in $Q^2$ 
(\mbox{$5 < Q^2 < 10\GeVSq$}, \mbox{$10 < Q^2 < 20\GeVSq$}
 and \mbox{$20 < Q^2 < 100\GeVSq$}).
%%%%%%%%%%%%%%%%%%%%%%%%%%%%%%%%%%%%%%%%%%%%%%%%%%%%%%%%%%%%%%%%
%%     LIMIT PLOTS   
%%%%%%%%%%%%%%%%%%%%%%%%%%%%%%%%%%%%%%%%%%%%%%%%%%%%%%%%%%%%%%%%
%%%%%%%%%%%%%%%%%%%%%%%%%%%%%%%%%%%%%%%%%%%%%%%%%%%%%%%%%%%%%%%%%%%%%%%%%%%
%mass spectra for 3 different Q2 bins, H1 dedx
%%%%%%%%%%%%%%%%%%%%%%%%%%%%%%%%%%%%%%%%%%%%%%%%%%%%%%%%%%%%%%%%%%%%%%%%%%%
%%%%%%%%%%%%%%%%%%%%%%%%%%%%%%%%%%%%%%%%%%%%%%%%%%%%%%%%%%%%%%%%%%%%%%%%%%%
The shape of the invariant mass distributions is found to be reproduced
by a background Monte Carlo simulation of inclusive DIS events using
the DJANGOH event generator\cite{django} and the H1 detector simulation
based on GEANT.
A fit of 
an empirical background function of the form 
\begin{equation}
f(M_{K^{0}_{s}p})=\alpha \cdot (M_{K^{0}_{s}p} - M_{thr})^{\beta} \cdot
\exp\{ -( M_{K^{0}_{s}p } - M_{thr}) \cdot \gamma \} 
\label{bgrfit}
\end{equation} 
is performed to the data, 
where $M_{thr}= M_{K^0_s} + M_p$ ($M_p$ being the proton mass)
and $\alpha$, $\beta$ and $\gamma$ 
are free parameters determined for each $Q^2$ interval independently.
The data are well described by this phenomenological function.
No narrow resonance is observed in any of the $Q^2$ bins. 
%%%%%%%%%%%%%%%%%%%%%%%%%%%%%%%%%%%%%%%%%%%%%%%%%%%%%%%%%%%%%%%%%%%%%%%%%%
%% Limit calculation
%%%%%%%%%%%%%%%%%%%%%%%%%%%%%%%%%%%%%%%%%%%%%%%%%%%%%%%%%%%%%%%%%%%%%%%%%%
The $M_{\ksf p}$ distribution is therefore used to set upper limits 
on the \thpl production cross section,
$$\sigmathf \equiv (\sigma(ep\ra e\thplf X)+ \sigma(ep\ra e\thmif X))\times BR( \thplf \ra  \knullf p)\quad .$$
Since the mass of the \thpl candidate is experimentally not well 
established, mass dependent limits are derived in the range
from 1.48 to 1.7$\GeV$.
%%
% For a given \thpl mass, $M_{\thplf}$, the hypothesis that a signal would
% lead to a number $N(M_{\thplf})$ of $\ksf p$ combinations
% is confronted with the data.
% This number is given by
For a given \thpl mass, $M_{\thplf}$, the expected number of
selected $\ksf p$ combinations due to \thpl production 
is related to $\sigmathf$ via
\begin{equation}
N(M_{\thplf})  = \sigmathf \cdot
  \displaystyle {\cal{L}} \cdot \epsilon_{DIS}
  \cdot\epsilon_{\thplf}(M_{\Theta^{+}}) 
  \cdot  BR(\knullf \ra \pi^+ \pi^-) \quad,
\label{nul}
\end{equation}
where 
${\cal{L}}$ is the integrated luminosity, 
$\epsilon_{DIS}$ is the acceptance  of the inclusive DIS event 
selection
and $\epsilon_{\thplf}(M_{\Theta^{+}})$ is 
the acceptance of the \thpl selection.  
The cross section $\sigmathf$ is integrated over the 
visible kinematic range studied, which is given by
\mbox{$\pt(\ksf p)>0.5\GeV$}, \mbox{$|\eta(\ksf p)| < 1.5$}, 
$0.1 < y < 0.6$ and the respective $Q^2$ bin. 
The branching ratio  for the transition of
\knull to \ks and its decay into charged pions 
is $BR(\knullf \ra \pi^+ \pi^-)=$ $BR(\knullf \ra \ksf)$ $\times$
$BR(\ksf \ra \pi^+ \pi^-)=$ \mbox{$0.5 \times (0.6895 \pm 0.0014)$ \cite{pdg}}. 
% No correction factor for the branching fraction
% of the \thpl decaying to $\knullf p$ is applied.

%%
An upper limit at the 95\% confidence level (C.L.) on $N(M_{\thplf})$,
$N_{UL}(M_{\thplf})$, is obtained from the observed, the background 
and the signal $M_{\ksf p}$ distributions in the mass range from 
1.45 to 1.8$\GeV$, using a modified 
frequentist approach based on likelihood ratios\cite{tjunk}. 
% This takes statistical and systematic uncertainties of the signal
% and the background number of combinations into account.
This takes into account statistical and systematic uncertainties of the signal
and the background number of $\ksf p$ combinations.
The $M_{\ksf p}$ distribution for signal combinations is taken to 
be a Gaussian with a mean $M_{\thplf}$ and a width corresponding
to the experimental mass resolution as obtained in the \thpl
Monte Carlo simulations. 
This width $\sigma(M_{\thplf})$ 
varies from 4.8 to 11.3$\MeV$ in the mass range 
from 1.48 to 1.7$\GeV$.
The background $M_{\ksf p}$ distribution is taken to be the
fitted function given by equation~(\ref{bgrfit}).
%% %%%%%%%%%%%%%%%%%%%%%%%%%%%%%%%%%%%%%%%%%%
%% systematic uncertainty on background events
%% %%%%%%%%%%%%%%%%%%%%%%%%%%%%%%%%%%%%%%%%%%

A systematic uncertainty on this background distribution is assessed 
by performing the fit under different assumptions:
using the background function (\ref{bgrfit}) in the full mass range, excluding 
a mass window  of $\pm 2 \sigma$ around the \thpl mass,
and also using the sum of the 
background function and a Gaussian with fixed mass $M_{\thplf}$ 
and width $\sigma(M_{\thplf})$ to account for a possible signal. 
The uncertainty of the number of background $\ksf p$ combinations
is estimated from the difference between the different fitting methods and
amounts to 2\%.

% The systematic uncertainty of the number $N(M_{\thplf})$ of observed
% $\ksf p$ combinations due to \thpl production comprises the following
% main contributions:
The systematic uncertainty of $N(M_{\thplf})$ comprises the following
main contributions:
%% %%%%%%%%%%%%%%%%%%%%%%%%%%%%%%%%%%%%%%%%%%
%% systematic uncertainty on signal events
%% %%%%%%%%%%%%%%%%%%%%%%%%%%%%%%%%%%%%%%%%%%
\begin{itemize}
%% %%%%%%%%%%%%%%%%%%%%%%%%%%%%%%%%%%%%%%%%%%
%% LUMI
%% %%%%%%%%%%%%%%%%%%%%%%%%%%%%%%%%%%%%%%%%%%
\item The measurement of the luminosity has an uncertainty of $1.5\%$.
%% %%%%%%%%%%%%%%%%%%%%%%%%%%%%%%%%%%%%%%%%%%
%% DIS, Trigger and weights 
%% %%%%%%%%%%%%%%%%%%%%%%%%%%%%%%%%%%%%%%%%%%
\item The  uncertainty of the inclusive DIS event selection,
$\epsilon_{DIS}$, is $6.5\%$, which is coming mainly 
from contributions due to the trigger efficiency (5\%), 
the SpaCal energy calibration (3\%),
remaining contamination from photoproduction background (2.5\%) and 
radiative corrections (1\%).
%% %%%%%%%%%%%%%%%%%%%%%%%%%%%%%%%%%%%%%%%%%%
%% Theta  + effi 
%% %%%%%%%%%%%%%%%%%%%%%%%%%%%%%%%%%%%%%%%%%%
\item The efficiency of the \thpl selection, 
$\epsilon_{\thplf}(M_{\Theta^{+}})$,
has an uncertainty of $8\%$ which comprises the uncertainty in modelling
track losses ($6\%$)
and the uncertainty in the efficiency of the \dedx selection ($5\%$). 
%% %%%%%%%%%%%%%%%%%%%%%%%%%%%%%%%%%%%%%%%%%%
%% model dependencies
%% %%%%%%%%%%%%%%%%%%%%%%%%%%%%%%%%%%%%%%%%%%
\item The Monte Carlo model used for correction
is based on the assumption that pentaquarks are produced 
with similar phase space distributions as strange baryons.
Since no established production mechanism for the
\thpl yet is known, production model dependent
uncertainties are not considered. 
Dependencies on the QCD models are estimated by comparing
the \thpl acceptances derived with the RAPGAP and the CASCADE \cite{cascade}
event generators, which incorporate different QCD evolution schemes. The 
difference is found to be small and negligible compared
with other sources of systematic uncertainties.
\end{itemize}

The contributions are added in quadrature and the resulting total 
systematic uncertainty of $N(M_{\thplf})$ is 11\%.
The uncertainty of the number of background $\ksf p$ combinations is the 
dominating systematic effect in the limit
calculation.

\begin{figure}[btp]
\begin{center}
\epsfig{figure=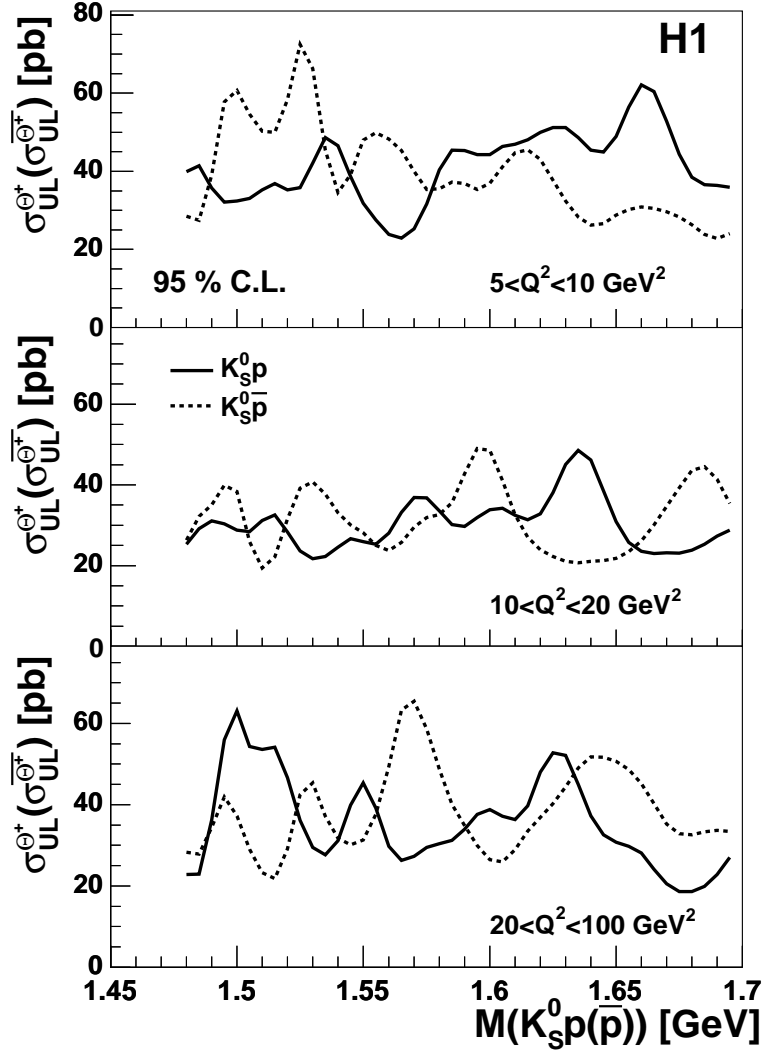,width=0.7\textwidth}
\caption{Upper limits on the cross section $\sigmaulthf$ (see text)
at 95\% confidence level in bins of $Q^2$ for $K^0_s p$ (full line) 
and $K^0_s \bar p$  (dashed line)
separately, integrated over the kinematic range
$\pt(K^0_s p)>0.5\GeV$, $|\eta(K^0_s p)|<1.5$ and $0.1 < y <0.6$.
\label{figlimitscharges}}
\end{center}
\end{figure}

The upper limit on the cross section, $\sigmaulthf$,
is then calculated from $N_{UL}(M_{\thplf})$ according to 
equation (\ref{nul}).
%%%%%%%%%%%%%%%%%%%%%%%%%%%%%%%%%%%%%%%%%%%%%%%%%%%%%%%%%%%%%%%%%%%%%%%%
The upper limits at 95\% C.L. 
are shown below the mass spectra of Fig.~\ref{figmkspandlimits} 
for the three different $Q^2$ bins. 
The  limits vary between \mbox{30 and 90\pb} for 
the different $Q^2$ bins and over the mass range from 1.48 to 1.7$\GeV$.
The invariant mass spectra of positive $\ksf p$ and negative $\ksf \bar p$
combinations are also studied separately. No narrow resonance is observed.
The corresponding upper limits 
for the $\thplf$ decaying to $\knullf p$ and its charge conjugate $\thmif$
decaying to $\overline{ \knullf }\bar p$,
shown in Fig.~\ref{figlimitscharges}, are found to be of comparable
size. 
The up- and downward fluctuations of the limits occur
at different masses for the different 
$Q^2$ bins and charges, which supports the hypothesis 
that the observed $\ksf p$ invariant mass distributions are
consistent with being due to combinatorial background only.

%%%%%%%%%%%%%%%%%%%%%%%%%%%%%%%%%%%%%%%%%%%%%%%%%%%%%%%%%%%%%%%%%%%%%%%%%%%
%% limits protons/antiprotons
%%%%%%%%%%%%%%%%%%%%%%%%%%%%%%%%%%%%%%%%%%%%%%%%%%%%%%%%%%%%%%%%%%%%%%%%%%%
%%%%%%%%%%%%%%%%%%%%%%%%%%%%%%%%%%%%%%%%%%%%%%%%%%%%%%%%%%%%%%%%%%%%%%%%%%%
% low momentum dedx selection: mass spectra 
%%%%%%%%%%%%%%%%%%%%%%%%%%%%%%%%%%%%%%%%%%%%%%%%%%%%%%%%%%%%%%%%%%%%%%%%%%%
\begin{figure}[bth]
\begin{center}
\epsfig{figure=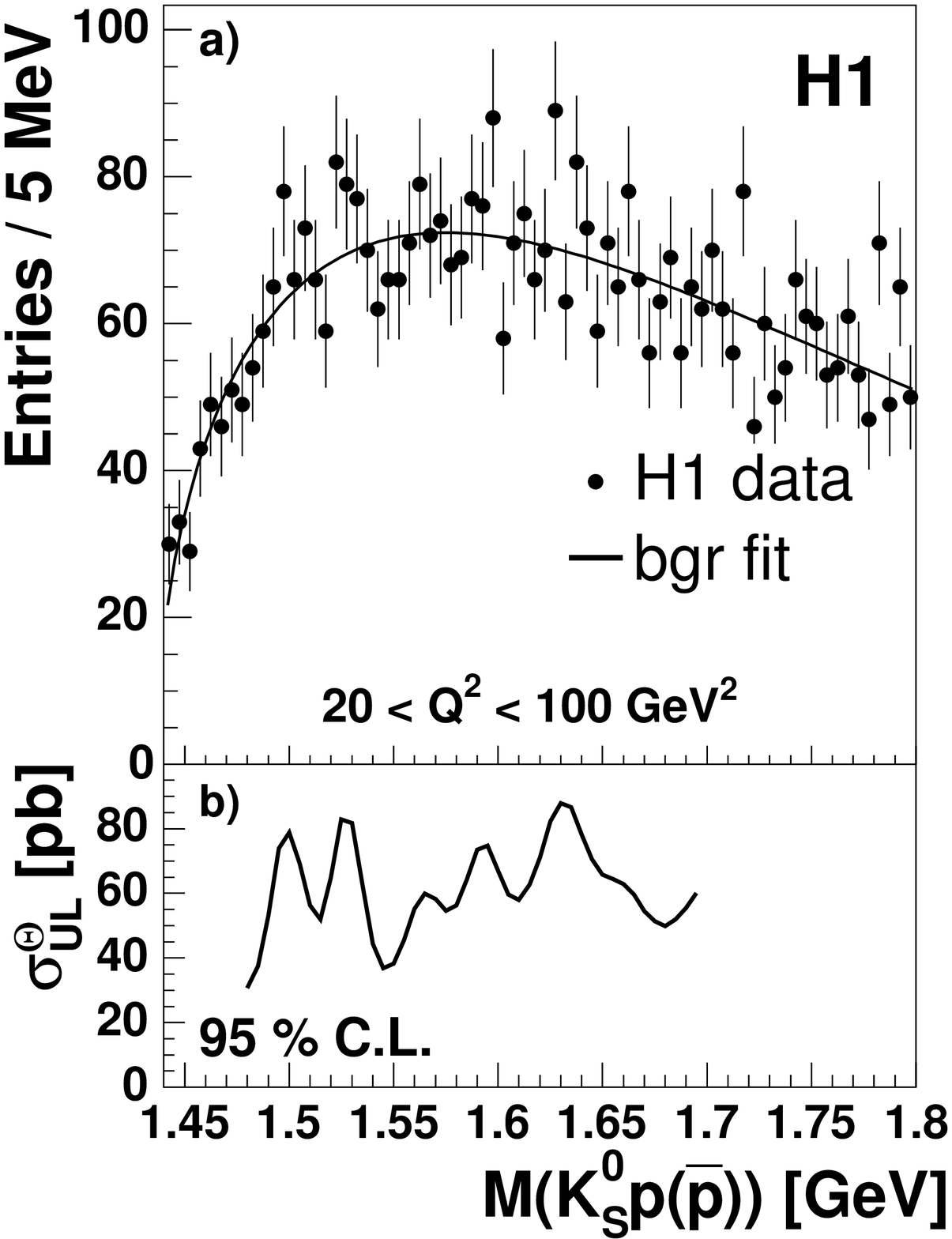,width=0.49\textwidth}
\epsfig{figure=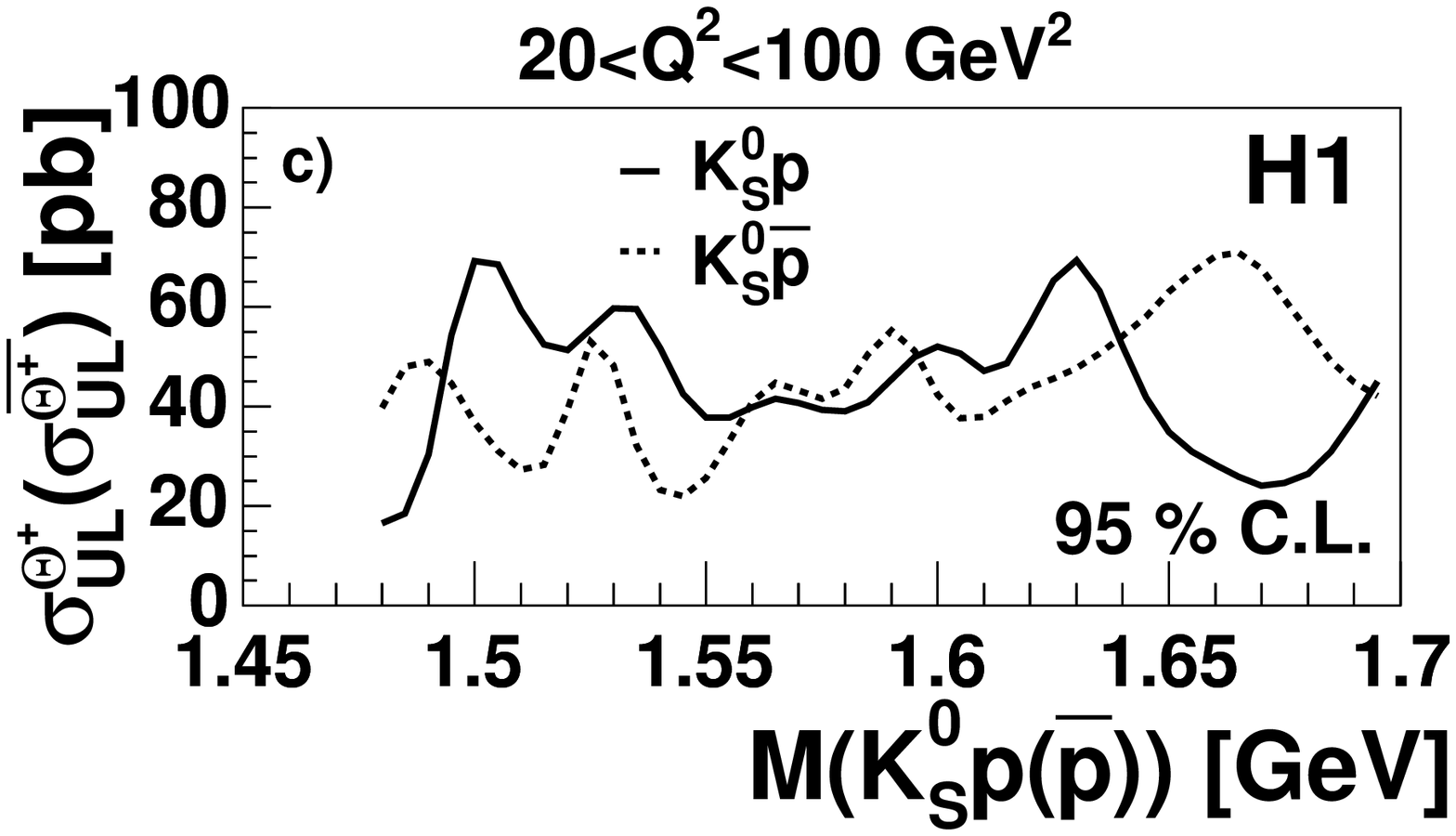,width=0.5\textwidth}
\caption{
a) Invariant $K^0_s p(\bar p)$ mass spectra for 
$20 < Q^2 < 100\GeVSq$ for proton candidates with momenta below 1.5$\GeV$ 
and b) upper limits on the cross section $\sigmaulthf$ (see text)
at 95\% confidence level for $20 < Q^2 <100\GeVSq$ integrated
over the kinematic range
$\pt(K^0_s p)>0.5\GeV$, $|\eta(K^0_s p)|<1.5$ and $0.1 < y <0.6$
using the low momentum proton selection for all $\ksf p$ combinations  
and c) for $K^0_s p$ (full line) and
$K^0_s \bar p$ (dashed line) combinations separately.
\label{figmkspzeusandlimit}}
\end{center}
\end{figure}
%%%%%%%%%%%%%%%%%%%%%%%%%%%%%%%%%%%%%%%%%%%%%%%%%%%%%%%%%%%%%%%%%%%
%% COMPARISON WITH ZEUS 
%%%%%%%%%%%%%%%%%%%%%%%%%%%%%%%%%%%%%%%%%%%%%%%%%%%%%%%%%%%%%%%%%%%
The ZEUS experiment has reported a positive  \thpl observation
at a mass of \mbox{1.522$\GeV$} in DIS for  \mbox{$Q^2 \ge 20\GeVSq$}
using a data sample corresponding to an integrated 
luminosity of 121 pb$^{-1}$\cite{zeusspq}.
The $\ksf p$ system was reconstructed using only protons having
a momentum below 1.5$\GeV$, while the requirements 
on the  transverse momenta and pseudorapidities of \ks and $\ksf p$ 
are the same as in the present analysis.
%%
% In order to compare the upper limits on the \thpl production
% more directly to the observation by the ZEUS experiment, 
% In order to search for \thpl production in a phase space similar
% to that in which the ZEUS experiment reports an observation,
The analysis described above is
repeated using only proton candidates with momenta below 1.5$\GeV$.
The resulting invariant $\ksf p(\bar p)$ mass spectra are shown in 
Fig.~\ref{figmkspzeusandlimit}a for 
$20 < Q^2 < 100\GeVSq$ and $0.1 < y <0.6$.
No significant pentaquark signal is observed for events in
the low momentum proton selection.
The upper limits $\sigmaulthf$
at 95\% C.L., derived from these mass spectra,
are shown in Fig.~\ref{figmkspzeusandlimit}b.
At a \thpl mass of \mbox{1.52$\GeV$} an upper limit on the cross section
of 72\pb at 95\% C.L. is found. 
The $\ksf p$  and $\ksf \bar p$  combinations do 
not yield any significant peak either. The corresponding upper limits
are also shown in Fig.~\ref{figmkspzeusandlimit}c.

%%%%%%%%%%%%%%%%%%%%%%%%%%%%%%%%%%%%%%%%%%%%%%%%%%%%%%%%%%%%%%%%%%%%
\section{Conclusions}

A search for the strange pentaquark \thpl
in deep inelastic $ep$ scattering is presented.
No signal for \thpl production  is observed in the decay mode 
$\thplf\ra\ksf p$ and $\thmif\ra\ksf \bar p$
for negative momentum transfers squared, $Q^2$, between 5 and $100\GeVSq$.
Assuming that pentaquarks are produced with 
similar kinematics as known strange baryons,
mass dependent upper limits at 95\% confidence level 
on the cross section 
$\sigma(ep\ra e\thplf X )\times 
BR(\thplf \ra  \knullf p)$
are derived in intervals of $Q^2$ and found to vary between 
\mbox{30 and 90\pb} over the mass range from 1.48 to \mbox{1.7$\GeV$}.

The analysis is repeated, restricted to large $Q^2$ and low proton momentum, 
a region in which the ZEUS collaboration observes evidence for a \thpl signal.
For this selection no signal is found either.

\newpage

%%%%%%%%%%%%%%%%%%%%%%%%%%%%%%%%%%%%%%%%%%%%%%%%%%%%%%%%%%%%
\section*{Acknowledgements}

We are grateful to the HERA machine group whose outstanding
efforts have made this experiment possible. 
We thank the engineers and technicians for their work in constructing and
maintaining the H1 detector, our funding agencies for 
financial support, the
DESY technical staff for continual assistance
and the DESY directorate for support and for the
hospitality which they extend to the non DESY 
members of the collaboration.

%%%%%%%%%%%%%%%%%%%%%%%%%%%%%%%%%%%%%%%%%%%%%%%%%%%%%%%%%%%%

\vspace{3cm}
%%%%%%%%%%%%%%%%%%%%%%%%%%%%%%%%%%%%%%%%%%%%%%%%%%%%%%%%%%%%%%%%%%%%%%%%%%%

\begin{thebibliography}{99}
%%
\bibitem{kenhicks} K.~Hicks, Prog Part. Nucl. Phys. {\bf 55} (2005) 647
[hep-ex/0504027].
\bibitem{theory}
D.~Diakonov, V.~Petrov and M.~Polyakov, Z. Phys.  A {\bf 359}, 305 (1997)
[hep-ph/9703373];\\
R.~L.~Jaffe and F.~Wilczek, Phys. Rev. Lett.  {\bf 91} (2003) 23
[hep-ph/0307341];\\
M.~Karliner and H.~Lipkin, Phys. Lett. {\bf B575} (2003) 249 [hep-ph/0307243];\\
for a review on pentaquark phenomenology see
R.~L.~Jaffe, Phys.~Rept. {\bf 409} (2005) 1 [hep-ph/0409065].
\bibitem{zeusspq} S.~Chekanov {\it et al.} [ZEUS Collaboration],
Phys. Lett. {\bf B591} (2004) 7 [hep-ex/0403051].
\bibitem{h1det} 
I. Abt {\it et al.} [H1 Collaboration], 
Nucl. Inst. Meth. A {\bf 386} (1997) 310; \\
I. Abt {\it et al.} [H1 Collaboration], Nucl. Inst. Meth. A {\bf 386} 
(1997) 348;\\
%%Spacal
R.~D.~Appuhn {\it et al.} [H1 SpaCal Group],
Nucl. Inst. Meth. A {\bf 386} (1997) 397;\\
%% what is this here???
C.~Adloff {\it et al.} [H1 Collaboration],
Eur. Phys. J. C {\bf 21} (2001) 33.
%%
\bibitem{steinhart} J. Steinhart, 
%`Die Messung des totalen 
%$c \bar{c}$-Photoproduktions-Wirkungsquerschnittes durch die
%Rekonstruktion von $\Lambda_c$ Baryonen unter
%Verwendung der verbesserten \dedx Teilchen\-identifikation am H1 
%Experiment bei HERA', 
Ph.D. thesis, 1999, Universit\"{a}t Hamburg 
(in German),
available from \\
http://www-h1.desy.de/publications/theses\_list.html.
%%
%%MC
\bibitem{rapgap} H.~Jung, Comput. Phys. Commun. {\bf 71} (1992) 15. 
%% no hep-ph
\bibitem{lund} B. Andersson, G. Gustafson, G. Ingelman and T. Sj\"ostrand, 
Phys. Rept. {\bf 97} (1983) 31.
%%no hep-ph
\bibitem{pythia} T. Sj\"{o}strand {\it et al.}, 
Comput. Phys. Commun. {\bf 135} (2001) 238 [hep-ph/0010017].
%%
\bibitem{strangenessdis}
M.~Derrick {\it et al.} [ZEUS Collaboration],
%  {\it Neutral strange particle production in deep inelastic
%    scattering at HERA},
Z. Phys. C {\bf 68} (1995) 29 [hep-ex/9505011];\\
S.~Aid {\it et al.} [H1 Collaboration],
%  {\it Strangeness Production in Deep Inelastic Positron-Proton
%    Scattering at HERA},
Nucl. Phys. B {\bf 480} (1996) 3 [hep-ex/9607010];\\
C. Risler, 
%`Die Produktion seltsamer neutraler Teilchen 
%in tiefinelastischer Streuung bei HERA`,
Ph.D. thesis, 2004, Universit\"{a}t
Hamburg (in German),
available from \\
http://www-h1.desy.de/publications/theses\_list.html.
%%
\bibitem{geant}  R. Brun {\it et al.}, GEANT3, Technical Report CERN-DD/EE/84-1,
CERN, 1987.
%% 
\bibitem{pdg} Particle Data Group, S. Eidelman {\it et al.}, 
Phys. Lett. {\bf B592} (2004) 1.
%%
\bibitem{django} K.~Charchula, G.A.~Schuler and H.~Spiesberger, Comput. Phys. Commun.
{\bf 81} (1994) 381.
\bibitem{tjunk} T.~Junk, Nucl. Inst. Meth. A {\bf 434} (1999) 435 [hep-ex/9902006].
%%
\bibitem{cascade} H.~Jung and G.P.~Salam, Eur. Phys. J. C {\bf 19} (2001) 
351 [hep-ph/0012143];\\
H.~Jung, Comput. Phys. Commun. {\bf 143} (2002) 100 [hep-ph/0109102].
%%
\end{thebibliography}
\end{document}